\documentclass{aa}
\usepackage{graphicx}
\usepackage{multirow}
\usepackage{txfonts}
\usepackage{floatflt}
\usepackage[authoryear]{natbib}       
\bibpunct{(}{)}{;}{a}{}{,}
\newcommand{\kms}{km~s$^{-1}$}
\newcommand{\KI}{\ion{K}{I}}
\newcommand{\NaI}{\ion{Na}{I}}
\newcommand{\CaI}{\ion{Ca}{I}}
\newcommand{\HI}{\ion{H}{I}}
\newcommand{\RbI}{\ion{Rb}{I}}
\begin{document}
\title{V4332 Sagittarii: A circumstellar disc obscuring the main object
 \thanks{Based in part on data collected at
    Subaru Telescope, which is operated by the National Astronomical
    Observatory of Japan.}}
\author{T. Kami\'nski 
\and M. Schmidt
\and R. Tylenda}  

\offprints{T. Kami\'{n}ski}
\institute{\center Department for Astrophysics, N. Copernicus
            Astronomical Center, 		  Rabia\'{n}ska 8,
            87-100 Toru\'{n}, Poland\\ 
            \email{tomkam, schmidt, tylenda@ncac.torun.pl}}
\date{Received; accepted}
\abstract
{V4332 Sgr experienced an outburst in 1994 whose
observational characteristics in many
  respects resemble those of the eruption of V838~Mon in 2002. It has been proposed
  that these objects erupted because of a stellar-merger event.}
{Our aim is to derive, from  observational data, information on the present
(10-15 yrs after the outburst) nature and structure of the object.} 
{We present and analyse a high-resolution ($R\approx21\,000$) spectrum of V4332 Sgr
  obtained with the Subaru Telescope in June~2009. Various components
(stellar-like continuum, atomic emission lines, molecular bands in emission)
in the spectrum are analysed and discussed. We also investigate 
a global spectral energy
  distribution (SED) of the object mostly derived from broadband
  optical and infrared photometry.}
{The observed continuum resembles that of an $\sim$M6 giant. The emission
features (atomic and molecular) are most probably produced by radiative
pumping. The observed strengths of the emission features strongly suggest
that we only observe a small part of the radiation of the main object
responsible for pumping the emission features. An infrared component seen
in the observed SED,  which can be roughly approximated by two
blackbodies of $\sim$950 and $\sim$200~K, is $\sim$50 times brighter than 
the M6 stellar component seen in the optical. 
This further supports the idea that the main object is
mostly obscured for us.}
{The main object in V4332~Sgr, an $\sim$M6 (super)giant, is surrounded by a
circumstellar disc, which is seen almost edge-on so the central star is
obscured. The observed M6 spectrum probably results from scattering the
central star spectrum on dust grains at the outer edge of the disc.}

        
\titlerunning{A disc in V4332 Sgr}
\authorrunning{Kami\'nski et al.}
\maketitle
\section{Introduction  \label{intro}}
V4332 Sgr is an unusual object, whose nova-like eruption was observed in 1994
\citep{martini}. Discovered as a K-type (super)giant, it showed a rapid
cooling on a time scale of months and declined as a very cool M-type
(super)giant. This evolution suggests that V4332~Sgr has
a similar nature to V838~Mon, which erupted in 2002. 
As discussed by \cite{ts06}, neither the
classical nova mechanism nor the He-shell flash model can explain the
observed outbursts of these objects. At present the stellar
collision-merger scenario proposed in \cite{st03} and
further developed in \cite{ts06} is the most promising
hypothesis for explaining the nature of these eruptions. 
 
Although it is now more than a decade after the outburst of V4332~Sgr, it
still shows very unusual observational characteristics. In the optical and
near-IR,
the object displays a very unique emission-line spectrum superimposed on a weak
continuum resembling a photospheric spectrum of a cool star
\citep{baner3,baner4,tyl,kimes}.
The emission spectrum shows very low excitation and is composed
 of lines from neutral elements (e.g. KI,
NaI, CaI) and molecular bands of TiO, ScO, AlO. 
Excitation temperatures derived from
analyses of the emission spectrum range
in 200--800\,K. The origin of this spectrum is not clear. 

V4332 Sgr is very bright in the infrared. Spectroscopic studies showed
strong absorption features of water ice at 3.05 and 6~$\mu$m, of silicates and
alumina at $\sim$10~$\mu$m, and possibly of methane ice at 7.7~$\mu$m, as
well as the fundamental band of CO in emission at 4.67~$\mu$m
\citep{banerCO,banerSpitzer}. Early studies of the spectral energy
distribution showed that, apart from the stellar-like component
dominating the observed continuum in the optical, the object displays a
bright infrared component with a characteristic temperature of $\sim$750~K and a
luminosity at least 15 times higher than that of the stellar one
\citep{tyl,baner4}. A possible circumstellar disc or ring orbiting the
central object was discussed in several studies \citep{baner4,banerCO,tyl}.
V4332~Sgr remained relatively constant in the optical brightness between
2002 and 2005. In 2006--2007, the object slowly faded by $\sim$1.5 mag in the $B$,
$V$, and $R$ bands, remaining almost unchanged in the $I$ band\footnote{see
the web page of V. Gornaskij: http://jet.sao.ru/$\sim$goray/v4332sgr.htm}.

In this paper we further investigate the nature
of V4332~Sgr. Our study is based on new high-dispersion spectrosopic observations
in June~2009 with the Subaru Telescope. We also discuss the global
spectral energy distribution (SED) of the object derived from different
measurements done in the optical, near-IR, and far-IR. We show that the
emission spectrum is most likely produced by radiative pumping. From our analysis of
the optical photospheric spectrum, emission features, and the SED, we
conclude that the central object in V4332~Sgr, most probably a cool
(super)giant, is hidden in a circumstellar disc seen almost edge-on.

\section{Observations}

\subsection{Optical spectroscopy with Subaru/HDS}
High-resolution spectroscopy of V4332~Sgr was obtained on 2009 June 16
with the High Dispersion Spectrograph \citep[HDS,][]{hds} on the 8.2--m
Subaru Telescope. Observations were performed in the standard {\it StdRb}
setting with two different echelle angles, resulting in a combined 
spectrum covering the range 5287--8070~\AA\ at a nominal resolution of
$R \simeq 21\,000$. Owing to the inter-chip gap and some
irregularities of the instrument's CCDs, the spectrum is missing
regions 6640--6708~\AA\ and 7340--7360~\AA, and has a few other minor
gaps. Five individual exposures were obtained with a total time of 100
min. 

Data were reduced with IRAF\footnote{IRAF is distributed by the
  National Optical Astronomy Observatories,
which are operated by the Association of Universities for Research
in Astronomy, Inc., under cooperative agreement with the National 
Science Foundation.} using standard procedures for
echelle spectroscopy \citep{irafech}. 
All the spectra were
wavelength-calibrated using a Th-Ar spectral lamp
and flux-calibrated using our observations 
of HR\,7596 and spectrophotometric fluxes from \cite{ham1,ham2}. 
We assess an overall accuracy of the relative flux calibration at $\sim15$\%\
and that of the wavelength calibration at 0.01~\AA.

\subsection{NIR photometry}\label{tcs}
Photometric $JHK_S$ observations of V4332 Sgr were carried out on 2009
May 18 with the CAIN-3 camera installed 
on the 1.52 m Carlos S\'{a}nchez Telescope (Observatorio del
Teide). Observations were performed in a sequence of exposures 4$\times$10
s, 4$\times$7 s, and 12$\times$3~s for the $JHK_S$ filters,
respectively. 
Data were reduced with the CAIN data reduction
scripts\footnote{http://www.iac.es/telescopes/cain/reduc/caindr.html} 
developed in IRAF by the Instituto de Astrof\'{i}sica de
Canarias. 
The data were calibrated to standard magnitudes by using
photometry from the All-Sky Point Source Catalog of the Two Micron All
Sky Survey (2MASS). 
The averaged magnitudes for V4332 Sgr are 
$J$=12.60$\pm$0.16, $H$=10.79$\pm$0.15, and $K_S$=9.12$\pm$0.15. 

\subsection{Submillimeter observations of CO(3--2)}\label{co_obs}
V4332~Sgr was also observed in the $^{12}$CO(3--2) (345.796 GHz) line
on 2009 May 14 using the James Clark Maxwell
Telescope\footnote{The James Clerk Maxwell Telescope is operated by The
  Joint Astronomy Centre on behalf of the Science and Technology
  Facilities Council of the United Kingdom, the Netherlands
  Organisation for Scientific Research, and the National Research
  Council of Canada.} (under the programme S09AI01). The HARP array
\citep{harp} was
used in the conventional stare mode.
The telescope's half-power beamwidth at the observed
frequency is about 15\arcsec. The observing method was beam-switching
with a chop
throw of 60\arcsec\ in azimuth. The ACSIS autocorrelator was used as a back-end
providing a spectral resolution of 0.42~km~s$^{-1}$. The total 
on-source integration time was 25~min and the typical system temperature
during observations was of $T_{\rm sys}=402$~K. 

We did not detect any emission in the LSR velocity range between
--500~km~s$^{-1}$ and 400~km~s$^{-1}$ for all of the HARP
receptors. The upper limit on the CO(3--2) emission at the position of
V4332~Sgr is 3$\sigma=78.7$~mK (per 0.42~km~s$^{-1}$ channel) on the
antenna's
temperature scale. This corresponds to 141~mK on the scale of the main beam
temperature (T$_{\rm mb}$), if the main beam efficiency of
$\eta_{\rm mb}=0.56$ is assumed.

\section{The optical spectrum of V4332 Sgr in 2009}

\begin{figure*}
\centering
   \includegraphics[height=\hsize,angle=270]{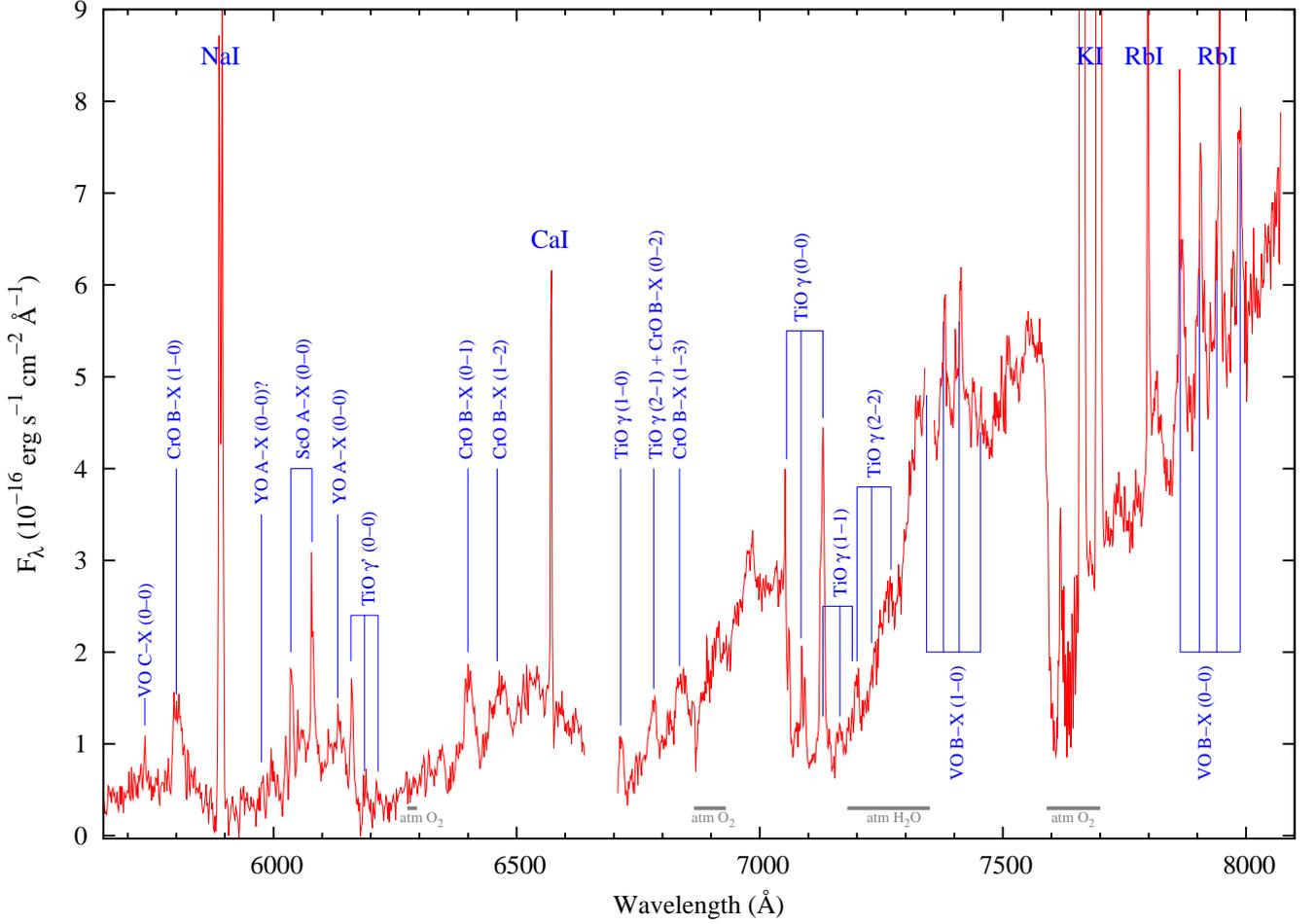}
   \caption{The best part of the spectrum of V4332 Sgr
   obtained with Subaru/HDS in June 2009 (red). The spectrum has been
   smoothed and dereddened with $E_{B-V}=0.32$ and $R_V=3.1$ (see text
   for details). The strongest emission features are
   identified.
   Parts affected by telluric absorption bands are indicated by
   horizontal bars at the bottom.} 
 \label{fig_spec}
\end{figure*}

The Subaru/HDS spectrum of V4332 Sgr appears similar to spectra that
have been reported for the post-outburst evolution of this object to
date \citep{baner4,tyl,kimes}. Strong lines of neutral atoms and molecular bands, all
seen {\it in emission}, are displayed on a cold M-type
photospheric spectrum. The best part of our spectrum is shown in 
Fig.~\ref{fig_spec}. Because the object is very faint, the raw spectrum
has a very poor signal-to-noise ratio (S/N) in the original resolution;
the average S/N per pixel in the continuum is of about 0.7 in the
5500~\AA\ region and $\sim$7 near 8000~\AA.  To identify
photospheric features, the spectrum was smoothed applying boxcar smoothing
with a box-size of 21 pixels. 
Fortunately,
most of the emission features are very strong, and they can be
studied in detail even at the original resolution. The photospheric
spectrum, and the atomic and molecular emission features,
as well as interstellar features, are all presented in detail in the following sections.    

\subsection{Photospheric spectrum  \label{sp_phot}}

\begin{figure}
\centering
   \includegraphics[height=\hsize,angle=270]{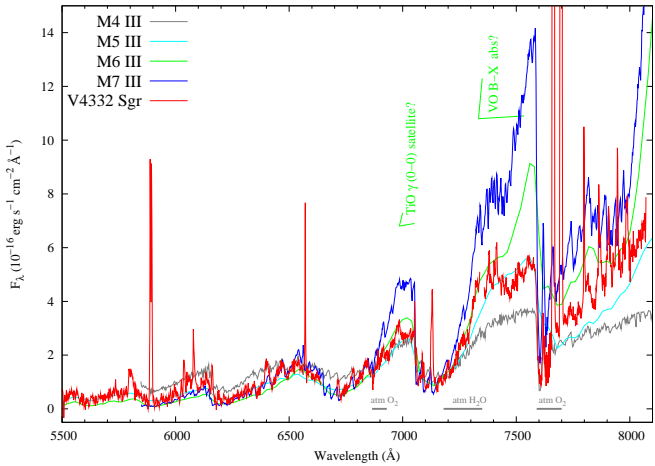}
   \caption{The 2009 spectrum of V4332 Sgr (red - the same as in
   Fig.~\ref{fig_spec}) compared to spectra of
   M-type giants taken from \cite{uves}$^5$ and \cite{gs}. 
  The regions suspected of being affected by some ``extra''
   molecular absorption are indicated by vertical green marks.}
 \label{fig_spec_2}
\end{figure}

A photospheric spectrum, characteristic of a late type star, can be easily
recognized in Fig.~\ref{fig_spec}. When compared with earlier spectroscopic
observations \citep{baner4,tyl,kimes}, the present spectrum displays much
more pronounced wide and deep absorption bands in the red part. This
obviously suggests a later spectral type and a lower effective temperature.
We constrained the spectral type of the present photospheric
spectrum.
The Subaru/HDS spectrum, corrected for interstellar extinction with $E_{B-V}=0.32$
(see Sect.~\ref{redden}), was compared
to a grid of observed spectra of late type giants with luminosity classes
III--II, taken from \cite{uves}\footnote{see also 
http://www.sc.eso.org/santiago/uvespop/field\_stars\_uptonow.html} 
and \cite{gs}. 
The result is shown in
Fig.~\ref{fig_spec_2}, where the giant (luminosity class III) spectra of 
types M4, M5, M6, and M7 are plotted against the spectrum of V4332~Sgr.
From this comparison we can
classify the V4332~Sgr spectrum as M5--6.
This can be compared with spectral
types K8--M3 derived from observational data in 2003 \citep{tyl,kimes}. The
decline in the spectral type most probably took place in 2006--2007 when the
object was gradually fading in $BVR$, as noted in Sect.~\ref{intro}. 
The derived spectral type of the 2009 spectrum corresponds to an effective 
temperature of $T_{\rm eff}\simeq3300$~K \citep{mgiants}. We have also
fitted standard photometric spectra of luminosity class III to 
the $BVR_cI_cJ$ photometric data
presented in Sect.~\ref{sed}. \citep[We used the same procedure as in][]{tyl} The
best fit gives a spectral type of M6.2 and $T_{\rm eff}\simeq3200$~K. 

We were not able to constrain the luminosity class of the
object, except that it is a giant.
There are no clear signs of absorption features of CaH,
characteristic of spectra of late-type dwarfs. 

The observed spectrum of V4332 Sgr differs
from any standard spectrum of a late-type giant. In addition
to the obvious presence of emission features, we noticed
{\it extra absorption bands} not observed (or being much
weaker) in stars of spectral types close to M6\,III. 
Most striking is the B--X absorption band of VO
in the range $\sim$7334--7534~\AA, which is more characteristic of
giants earlier than M6. Some
extra absorption can also be found in the range $\sim$6985--7050~\AA;
it can be assigned to the satellite branch of TiO $\gamma$ (0,0)
\citep[cf.][]{KST09}, but this identification is very tentative. The
regions contaminated by the extra
absorption features are indicated in  Fig.~\ref{fig_spec_2}. 

A possible explanation for the difference between the observed spectrum of
V4332~Sgr and the standard spectra of giants could be
the peculiarities in the object's metallicity. We compared the Subaru/HDS spectrum
with a grid of MARCS spectra \citep{marcs} with $T_{\rm eff}=3200$~K, 
$\log g=0.0$, M\,=\,1\,M$_{\sun}$, standard chemical composition, and
metallicity in the range  $-2 \leq [$Fe/H$] \leq +0.5$
(see the table on http://marcs.astro.uu.se). The shape of
the observed spectrum is best represented by the model spectrum with
[Fe/H]\,$=-1.5$. The fit is still not perfect but no extra absorption is
needed in the 7500~\AA\ region; 
however, \cite{martini}
obtained fits to the spectra observed during the 1994 outburst using models
with solar abundances. Our suggestion of a low metallicity in V4332~Sgr is
very tentative. The object is not a typical giant, so peculiarities in the
spectrum are expected. Besides, as discussed in Sect.~\ref{sed}, we probably
do not observe the object directly, but rather a part of its spectrum that
is scattered on
a disc matter. Scattering can affect the shape of the spectrum.

We did not find any atomic absorption features that could be identified as
photospheric lines. A photospheric absorption line of
\CaI\,6572\,\AA, relatively strong in late type giants, may be
the cause of the irregular shape of the observed \CaI\ emission profile,
as discussed in Sect.~\ref{emlines}.
No photospheric signatures were found within the profiles
of the \NaI\ and \KI\ doublets. 

The lack of atomic absorption lines in the photospheric spectrum does not
allow us to derive a reliable value of the radial velocity of the object. As
can be seen from Fig.~\ref{fig_spec}, most of the photospheric absorption bands
coincide with narrow components in emission. However, in a few cases
relatively clean band-heads without emission component can be seen and their
position measured. This is the case for
the following TiO $\gamma$ bands: (3,2) R$_{1}$ (6849$\pm$1.0 \AA), 
(4,3) R$_1$ (6917.6$\pm$0.5 \AA), (5,4) R$_1$ (6987.4$\pm$1.0 \AA),
(0,0) $^S$R$_{21}$ (7036.7$\pm$0.5 \AA), and (3,4) R$_3$
(7818.6$\pm$0.5\,\AA).
These measurements give rough estimates of the (heliocentric) radial velocity of the
object, i.e.,  $-$40, $-$56, $-$45, $-$67,
and $-$60~\kms, respectively. These values imply
a heliocentric radial velocity
of the source of the photospheric spectrum of $-56 \pm 16$~\kms.

\subsection{Atomic emission lines \label{emlines}}

\begin{table*}
\begin{minipage}[t]{\hsize}
\caption{Measurements for the resonance emission lines seen in the Subaru
spectrum
  of V4332~Sgr.}
\label{tab_atomic}
\centering
\renewcommand{\footnoterule}{} 
\begin{tabular}{cccccrcccc}
\hline
\hline
$\lambda_{\rm lab}$ (air)&
$\lambda_{\rm obs}$ &
Ion&
$V_{\rm h}$&
FWHM\footnote{from a fit of a Lorentzian profile; full width at
half-maximum is not corrected for the instrumental profile 
(FWHM$\approx$ 0.15~\AA)}&
FWHM$^{a}$&
Flux&
Flux clean\footnote{integrated flux with interstellar and strongest
telluric features removed from the profile}&
EW\\[0pt]     
[\AA]&[\AA]& &[km s$^{-1}$]&[\AA]&[km s$^{-1}$]
&[erg s$^{-1}$ cm$^{-2}$]&[erg s$^{-1}$ cm$^{-2}$]&[\AA]\\[2pt]
\hline
&&&&&\\[-8pt]
5889.950& 5888.82$\pm$0.14&\ion{Na}{I} D$_2$& --57.5$\pm$7.2&
2.72$\pm$0.05&138.5& (1.81$\pm$0.01)e--15\footnote{results from a
deblending procedure with a double Lorentzian profile; simple integral
for both doublet lines gives summary flux of (3.26$\pm$0.06)e--15 erg
s$^{-1}$ cm$^{-2}$ and cleaned flux of
(3.48$\pm$0.06)e--15 erg s$^{-1}$ cm$^{-2}$}& (2.21$\pm$0.06)e--15$^c$&\\
5895.924& 5894.77$\pm$0.07&\ion{Na}{I} D$_1$& --58.7$\pm$3.6&
2.68$\pm$0.06&136.3& (2.00$\pm$0.01)e--15$^c$& (2.22$\pm$0.07)e--15$^c$&\\
6572.779& 6571.12$\pm$0.02&\ion{Ca}{I}      & --75.7$\pm$0.7&
1.21$\pm$0.13& 55.2& (7.34$\pm$0.43)e--16&         &  ~~~9.5$\pm$1.3~\\
7664.911& 7663.13$\pm$0.03&\ion{K}{I}       & --69.7$\pm$1.1&
3.54$\pm$0.02&138.5& (5.73$\pm$0.02)e--14&
(6.00$\pm$0.02)e--14$^d$&285.4$\pm$33.1\\
7698.974& 7697.22$\pm$0.01&\ion{K}{I}       & --68.3$\pm$2.9&
3.36$\pm$0.04&130.9& (4.89$\pm$0.01)e--14$^d$& (4.83$\pm$0.02)e--14$^d$&
267.2$\pm$17.7\\
7800.268& 7798.52$\pm$0.01&\ion{Rb}{I}      & --67.2$\pm$0.2&
1.58$\pm$0.31& 60.7& (1.49$\pm$0.08)e--15~&          & ~~~6.0$\pm$0.6~\\
7947.603& 7945.98$\pm$0.03&\ion{Rb}{I}      & --61.2$\pm$1.3&
1.01$\pm$0.08\footnote{line blends with a TiO band; width and flux
measurements uncertain}
                                                 & 38.1$^e$ &
(3.54$\pm$0.13)e--16$^e$&            &   ~~~0.85$\pm$0.1$^e$\\
\hline
\end{tabular}
\end{minipage}
\end{table*}

\begin{figure}
\centering
  \resizebox{\hsize}{!}{\includegraphics{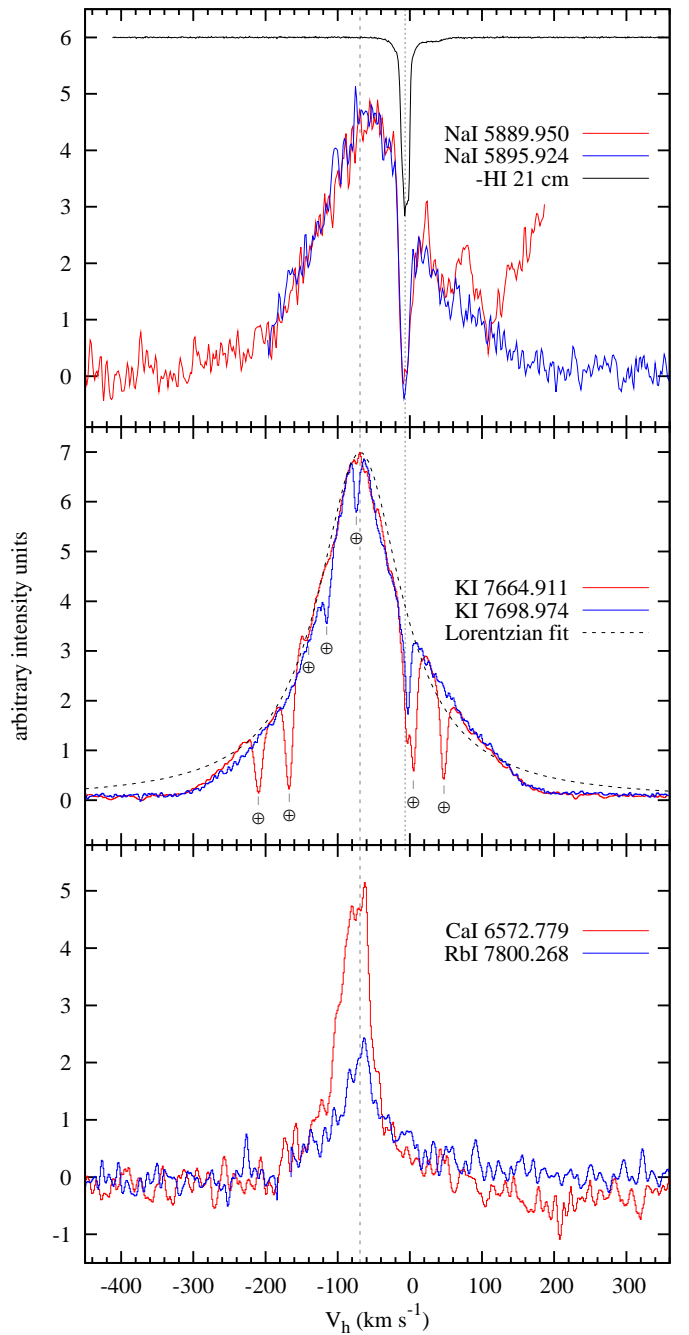}}
   \caption{Profiles of the strong resonance lines in the
     spectrum of V4332 Sgr. Spectra are rescaled. The vertical
     lines mark the centre of the \KI\ lines at --69 \kms\ and the
     centre of interstellar absorption features of \NaI\ at --7.2 \kms. {\bf
       Top:} The \NaI\ profiles, clipped for clarity. The top
     spectrum shows the intensity-inverted profile of the \HI\ 21 cm
     line (see text). {\bf Middle:} The observed \KI\ lines and a theoretical
     Lorentzian profile. The strongest telluric absorption lines are 
     indicated. {\bf Bottom:} \CaI\ and \RbI\ lines, spectra slightly smoothed to
     reduce the noise.}
 \label{fig1}
\end{figure}

The strongest atomic emission lines seen in the Subaru/HDS spectrum
are listed in Table\,\ref{tab_atomic}. They are characterized in terms
of observed wavelength, heliocentric radial velocity (V$_h$), full width at
half-maximum (FWHM), integrated flux, and equivalent
width (EW). For the \KI\ and \NaI\ doublets, we also present `cleaned'
fluxes, which are fluxes measured for the lines after removing all
atmospheric and interstellar absorption features from
their profiles. The fluxes are not
corrected for interstellar reddening. We do not present any measurements of
EW for the \NaI\ doublet since the observed continuum in the region of these
lines is very faint. Some of
the strongest emission lines are indicated in the spectrogram presented in
Fig.\,\ref{fig_spec}. Their profiles are displayed in more detail in
Fig.~\ref{fig1}.

The profiles of the emission lines are triangular in shape and their cores
can be reproduced well by a Lorentzian profile (see
Fig.\,\ref{fig1}, middle panel); however, the observed wings are
significantly less extended than in the Lorentzian profile.
The emission
lines appear to be fairly symmetric. Only with the mirror profile-inverting procedure or
by over-plotting a theoretical profile on the observed ones can
some asymmetry be noticed (see Fig.\,\ref{fig1}). 
The \CaI\ line shows a clear sign of asymmetry, with a
stronger blueshifted half of the profile, but this may be an effect of a
non-flatness of the underlying continuum or a contamination by a
photospheric absorption line (if present).

\subsection{Molecular bands in emission \label{embands}}

All the identified molecular emission features are listed in
Table~\ref{TableMolec}. Columns
(1)--(2) of the table contain the observed and laboratory wavelengths of the feature.
Columns (3)--(4) show the name of the molecule and identification of the electronic
system, vibrational band, and the branch forming the head. Column (5)
presents the heliocentric radial velocities of those features for which a unique
assignment was reliable. Column (6) shows the integrated fluxes of the
emission lines (for which reliable measurements were possible), while column
(7) lists their equivalent widths.
Comments on the nature of the features are given in column (8), while the
sources of the laboratory wavelengths can be found in column (9).

The laboratory wavelengths refer to the band-head positions.
When a sharp band-head is recognized in the spectrum, 
the observed wavelength corresponds to the position of the band-head.
These values have been used when determining the radial velocities.
In the other cases, the observed wavelengths are the positions of apparent maxima 
of the emissions. In the case of
the VO B$^4\Pi$--X$^4\Sigma^{-}$ bands, the four emission features
correspond to the four sub-bands formed by the main branches concentrated around
the sub-band origin.

The emission fluxes were
measured by integrating the surface above the continuum. The continuum
position has usually been determined by an inspection of the VLT/UVES spectra 
of the M5~III, M6~III, and M7~III stars (HD~214952, HD~189124, 
and HD~18242, respectively) \citep{uves}. 
In rare cases of gaps in the comparison spectra,
a theoretical spectrum was synthesized to exclude
the possibility of strong underlying absorptions.
Large  errors in the flux measurements of the VO B$^4\Pi$--X$^4\Sigma^{-}$
(0,0) are mainly due to uncertainties in the underlying continuum.

A detailed analysis of the profiles of the bands is presented in
Sect.~\ref{embands_mod}

\begin{table*}
\begin{minipage}[t]{\linewidth}
\caption{Molecular emissions}
\label{TableMolec}
\renewcommand{\footnoterule}{}
\begin{tabular}{llclccclr}
\hline
$\lambda_{\rm{obs}}$ & $\lambda_{\rm{lab}}$ & Molecule & Identification &
 Velocity & Flux & EW & Comment & Reference \\[0pt]
 [\AA] & [\AA] & & &
 [km s$^{-1}$] & [$10^{-16}$ erg s$^{-1}$ cm$^{-2}$] & [\AA] & & \\
\hline
5449 & 5448.233   & TiO & $\alpha$ (0,1)  R$_2$ &
   &  2.8$\pm$0.5 & 38$\pm$22 & noisy &  1  \\
5474 & 5469.3     & VO &  C$^4\Sigma^{-}$--X$^4\Sigma^{-}$  (1,0)  R$_4$ &
   &  0.6$\pm$0.3 &  5$\pm$4  & weak  &  1  \\
5736 & 5736.703   & VO &  C$^4\Sigma^{-}$--X$^4\Sigma^{-}$  (0,0)  R$_4$ &
   & $\la 1.0$       &               & weak  &  1  \\
5795 & 5794.6\footnote{remaining four heads are at positions 
5796.0 (R$_2$), 5797.5 (R$_3$), 5801.0 R$_4$, and 5806.8 R$_5$
\citep{Rosen1970}}
 & CrO & B$^5\Pi$--X$^5\Pi$ (1,0) R$_1$ &
  $-$70$\pm$20\footnote{values determined from simulations of the rotational
contour}
 & 7.0$\pm$0.6 & 28$\pm$5  &  &  2  \\
6036 & 6036.17    & ScO & A$^2\Pi_{3/2}$--X$^2\Sigma^{+}$ (0,0) $^{\rm{R}}$Q$_{2G}$ &
   $-$75$\pm$10$^{\mathrm{b}}$ & 3.9$\pm$0.6 & 14.5$\pm$4.5 &  &  1  \\
6058 & 6051.8\footnote{remaining four heads are at positions 
  6053.3 (R$_2$), 6054.8 (R$_3$), 6058.5 (R$_4$), 6063.5 (R$_5$)
\citep{Rosen1970}}
 & CrO & B$^5\Pi$--X$^5\Pi$ (0,0) R$_1$ &
   & 3.5$\pm$1.0  & 12.5$\pm$3.5 & noisy &  2  \\
6078 & 6079.30  & ScO & A$^2\Pi_{1/2}$--X$^2\Sigma^{+}$ (0,0) $^{\rm{Q}}$Q$_{1G}$ &
   $-$75$\pm$10     &  7.0$\pm$0.6 & 21.0$\pm$5.0 &  &  1  \\
6133 & 6132.097 & YO  & A$^2\Pi_{1/2}$--X$^2\Sigma^{+}$ (0,0) $^{\rm{R}}$Q$_{21}$ &
   & 1.2$\pm$0.6 & 2.5$\pm$0.7 &  &  1  \\
6161 & 6158.52  & TiO & $\gamma$' (0,0) $R_{1}$ &
   & 4.5$\pm$0.8 & 31$\pm$8 &  &  1  \\
6190 & 6186.32  & TiO & $\gamma$' (0,0) $R_{2}$ &
   & 1.1$\pm$0.8 & 13$\pm$7 & noisy &  1  \\
6212 & 6214.93  & TiO & $\gamma$' (0,0) $R_{3}$ &
   & 0.8$\pm$0.8 & 5$\pm$5  & &  1  \\
6352 & 6351.29  & TiO & $\gamma$ (2,0) R$_{1}$  &
   & $<$0.7 & & &  1  \\
6400 & 6394.2\footnote{remaining four heads are at positions 
 6396.2 (R$_2$), 6397.8 (R$_3$), 6401.4 (R$_4$), 6407.7 (R$_5$)
\citep{Rosen1970}}
  & CrO &  B$^5\Pi$--X$^5\Pi$ (0,1) R$_1$ &
  $-$70$\pm$20$^{\mathrm{b}}$ & 4.8$\pm$0.7 & 10$\pm$6 &  &  2  \\
6418 & 6414.742  & TiO & $\gamma$ (3,1) R$_{1}$  & & & & uncertain &  1  \\
6449 & 6451.7\footnote{wavelengths of the band heads derived from
spectroscopic constants of \cite{Hocking}:
 6451.953 (R$_2$), 6455.191 (R$_3$), 6459.463 (R$_4$), 6465.369 (R$_5$)}
 & CrO & B$^5\Pi$--X$^5\Pi$ (1,2) R$_1$ &
   & $\la$1.0 &  & uncertain  &  5  \\
6478 & 6478.953  & TiO & $\gamma$  (4,2) R$_{1}$  &
   & $\la$1.5 & & weak &  1  \\
6717 & 6714.477 & TiO & $\gamma$  (1,0) R$_{1}$  &
   & 2.7$\pm$0.5 & 14$\pm$4 &  &  1  \\
6772 & 6772.3\footnote{remaining four heads are at positions 
 6774.2 (R$_2$), 6775.9 (R$_3$), 6779.6 (R$_4$), 6785.7 (R$_5$)
\citep{Rosen1970}}
  & CrO & B$^5\Pi$--X$^5\Pi$ (0,2) R$_1$ &
  & & & blend with TiO &  2  \\
6785 & 6781.815 & TiO & $\gamma$  (2,1) R$_1$ &
   & 3.1$\pm$0.5 & 8$\pm$3 & blend with CrO &   1  \\
6837 & 6836.5\footnote{wavelengths of the band heads extrapolated from
spectroscopic constants of \cite{Hocking}:
6836.6 (R$_2$), 6839.9 (R$_3$), 6844.5 (R$_4$), 6850.9 (R$_5$)}
   & CrO & B$^5\Pi$--X$^5\Pi$ (1,3) R$_1$ &
   & & & uncertain &  5  \\
7052.5     & 7054.256 & TiO & $\gamma$  (0,0) R$_{3}$ &
   $-$80$^{\mathrm{b}}$ &  6.6$\pm$1.5 & 14$\pm$4 & &  1  \\
7086.2     & 7087.566 & TiO & $\gamma$  (0,0) R$_{2}$ &
   $-$80$^{\mathrm{b}}$ &  5.3$\pm$0.9 & 14$\pm$5 & &  1  \\
7129.4     & 7125.510 & TiO & $\gamma$  (0,0) R$_{1}$ &
   $-$80$^{\mathrm{b}}$ & 17.6$\pm$1.2 & 44$\pm$4 & blend (1,1) R$_{3}$ &  1  \\
7165       & 7158.850 & TiO & $\gamma$  (1,1) R$_2$   &
   & 1.3$\pm$0.4 & 3$\pm$1 & noisy &  1  \\
7199.3     & 7196.373 & TiO & $\gamma$  (2,2) R$_3$   &
   & 2.8$\pm$0.7 & 5$\pm$1 & blend (1,1) R$_1$ &  3  \\
7269.3     & 7269.984 & TiO & $\gamma$ (2,2) R$_1$    &
   & $<$1.1 & & weak &  1  \\ 
7381       &          & VO & B$^4\Pi$--X$^4\Sigma^{-}$ (1,0) F$_2$--F$_2$ &
   & 10.0$\pm$0.7 & 5$\pm$2 &  & \\
7413       &          & VO & B$^4\Pi$--X$^4\Sigma^{-}$ (1,0) F$_3$--F$_3$ &
   & 10.5$\pm$1.0 & 4$\pm$1.5 &  & \\
7590.4     & 7589.285 & TiO & $\gamma$ (0,1) R$_{3}$  &
   & 6.5$\pm$3.0 & 6.5$\pm$3 & telluric O$_2$ &  3  \\
7740.7   & 7743.045 & TiO & $\gamma$ (2,3) R$_{3}$  &
   $-$91$\pm$39 & & & noisy &  3  \\
7826.3 & 7828.153 & TiO & $\gamma$ (2,3) R$_{1}$  &
   $-$71$\pm$11 & & & noisy & 4 \\
7864       &          & VO & B$^4\Pi$--X$^4\Sigma^{-}$ (0,0) F$_1$--F$_1$  &
   $-$75$\pm$5$^{\mathrm{b}}$ & 26$\pm$2 & 12$\pm$1 & & \\
7907       &          & VO & B$^4\Pi$--X$^4\Sigma^{-}$ (0,0) F$_2$--F$_2$  &
   $-$75$\pm$5$^{\mathrm{b}}$ & 27$\pm$8 & 11$\pm$1 & & \\
7946       &          & VO & B$^4\Pi$--X$^4\Sigma^{-}$ (0,0) F$_3$--F$_3$  &
 $-$75$\pm$5$^{\mathrm{b}}$ & 36$\pm$8  & 15$\pm$4 & blend with Rb I & \\
7987       &          & VO & B$^4\Pi$--X$^4\Sigma^{-}$ (0,0) F$_4$--F$_4$  &
 $-$75$\pm$5$^{\mathrm{b}}$ & 29$\pm$10 & 11$\pm$4  &  & \\
\hline
\end{tabular}

\begin{list}{}{}
\item{References: (1) see references in \cite{KST09}; 
(2) \cite{Rosen1970}; 
(3) \cite{Schwenke}; (4) \cite{Phillips1973};
(5) \cite{Hocking}}
\end{list}

\end{minipage}

\end{table*}

\subsection{Interstellar features \label{ism}}

The only features that can be ascribed to the interstellar medium (ISM) are the \NaI\
5889/5895 and \KI\ 7664/7698 doublets in absorption. No diffuse
interstellar bands were found in the spectrum.   
We determined radial velocities, FWHMs, and EWs of
the interstellar features by fitting Gaussian profiles. Before performing 
a fitting procedure, the appropriate regions of the spectrum
were normalized with high order polynomials to correct for the broad
emission features that form a steep baseline for the interstellar
features. The results of Gaussian fits are listed in Table \ref{tab_ISM}. The
errors of the EWs reflect only a statistical error,
which was determined from a series of independent measurements for
each line. The FWHMs given in Table~\ref{tab_ISM} are not corrected for the
instrumental broadening. The FWHM of the instrumental profile,
as measured from the narrowest telluric lines seen in the spectrum, is on
average 8.0~\kms, so the profiles of the interstellar features are
mostly resolved.

\begin{table}
\begin{minipage}[t]{\hsize}
\caption{Results of Gaussian fits to the ISM absorption features found in
  the Subaru/HDS spectrum of V4332~Sgr from June 2009.}
\label{tab_ISM}
\centering 
\renewcommand{\footnoterule}{} 
\begin{tabular}{ccccc}
\hline
\hline
$\lambda_{\rm obs}$&Ion&$V_{\rm h}$&FWHM&EW\\[0pt]      
[\AA]& &[km s$^{-1}$]&[km s$^{-1}$]&[\AA]\\[2pt]
\hline
&&&\\[-8pt]
5889.950& \ion{Na}{I} D$_2$&--7.13& 15.78&0.40 ($\pm$0.04)\\  
5895.924& \ion{Na}{I} D$_1$&--7.32& 13.73&0.34 ($\pm$0.02)\\  
7664.911& \ion{K}{I}      &~--3.95\footnote{feature blends with a telluric line; measurements uncertain}&~11.34$^a$&~0.18 ($\pm$0.03)$^a$\\ 
7698.974& \ion{K}{I}      & --3.27& 9.73&0.12 ($\pm$0.01)\\  
\hline
\end{tabular}
\end{minipage}
\end{table}
  
As can be seen in Table~\ref{tab_ISM}, there is a shift of about 4
\kms\ between the peaks of the \ion{Na}{I} and \ion{K}{I} lines. The
shift may be physical, but it can also be partially caused by the fact that the
\NaI\ absorption lines are saturated or close to saturation.

The interslellar absorption features seen in the spectrum of V4332~Sgr can be
compared to a profile of the \HI\ 21 cm emission line in the
direction of the star. From the 21~cm LAB Survey \citep{lab}, we
extracted a spectrum for a position nearest to the position of
V4332~Sgr, i.e. at $l=13 \fdg 50$, $b=-9 \fdg 50$ (V4332 Sgr: $l=13
\fdg 63$, $b=-9 \fdg 40$). The half-power beam-width of the LAB Survey
is 36\arcmin, so the position of V4332 Sgr is located close to the
centre of the beam for the extracted position. The profile of the \HI\
line is composed of several narrow components, tightly blending into
one complex profile.
The intensity-inverted and rescaled spectrum of
\HI\ is shown in Fig.~\ref{fig1} (top panel). A Gaussian fit to the overall
profile gives a central velocity of V$_h=-6$ km s$^{-1}$. The profiles
of the absorption lines of \KI\ and \NaI\ lie well within the
profile of \HI.  The interstellar absorption lines seen in the optical
spectrum indicate that the line of sight towards V4332 Sgr crosses
most, if not all, of the ISM in this direction. 
This conclusion agrees with an analysis of
extinction of V4332 Sgr in \cite{kimesConf}.  

\section{Reddening, distance, radial velocity \label{redden}}

Analyses of spectroscpic and photometric observations of V4332~Sgr during 
its outburst in 1994 have led
\cite{martini}, \cite{tyl}, and \cite{kimes} to conclude that the reddenning to 
the object is $E_{B-V}=0.32\pm0.02$, $E_{B-V}=0.32\pm0.10$, and
$E_{B-V}=0.37\pm0.07$, respectively.

The reddenning can be estimated from a relation between EWs of atomic 
interstellar absorption lines and  $E_{B-V}$. Using
the relation derived in \cite{ebv}, our measurement for the \ion{K}{I}
7699 line from Table~\ref{tab_ISM} results in $E_{B-V}=0.45\pm0.05$. 
We did not use the \ion{Na}{I}~D lines since they
reach zero flux in
our spectrum and are probably saturated.

We also used the Galactic Dust Extinction
Service\footnote{http://irsa.ipac.caltech.edu/applications/DUST/},
which estimates $E_{B-V}$ from the 100~$\mu$m dust
emission mapped by IRAS and COBE/DIRBE. This tool is based
on a method that gives good constraints on the extinction for
regions with low-to-moderate reddening. For the position of V4332~Sgr,
we get $E_{B-V}=0.344\pm0.008$.

An independent constraint on the reddening can be obtained from the radio
observations of the \HI\ 21~cm line. The integrated intensity of the
\HI\ line of 664.2~K\,\kms\ can be 
converted to a column density of 
$N({\rm \ion{H}{I}})=1.25\times10^{21}$~cm$^{-2}$ using a standard conversion
factor of $1.822\times10^{18}$~cm$^{-2}$~(K~\kms)$^{-1}$
\citep[e.g.,][]{tools}.  The value of the reddening can be found from
a relation given by \cite{bohlin}.
By neglecting a contribution from molecular hydrogen (our
observations in the CO rotational transition, see Sect.~\ref{co_obs}, show no molecular cloud
in the direction of V4332 Sgr)
we get $E_{B-V}=0.22\pm0.04$. 
In the present paper we adopt $E_{B-V}=0.32$ as a reasonable
compromise between the above estimates.

The distance to V4332~Sgr is unknown. \cite{martini} estimated that the
object is at $\sim 300$~pc. Assuming that the progenitor of V4332~Sgr was a
main sequence star, \cite{tyl} estimated a distance of $\sim 1.8$~kpc. A
lower limit to the distance can be obtained from the observed radial
velocities of the interstellar absorption lines.
A mean heliocentric radial velocity derived from the values listed in 
Table~\ref{tab_ISM} is $-5.5$~\kms. This corresponds 
to $V_{\rm LSR} = 6.5$~\kms.
After adopting a Galactic rotation curve of \cite{bb93}, we find that V4332~Sgr is
at a distance $\ga 1.0$~kpc. This excludes the low distance value of
\cite{martini}.

 As noted in \cite{tyl}, the radial velocity of V4332~Sgr shows that the
object does not follow the Galactic rotation.
The present analyses of the photospheric spectrum, atomic emission lines, and
molecular bands give $V_h = -56 \pm 16$\,km\,s$^{-1}$
(Sect.~\ref{sp_phot}), $-65 \pm 7$\,km\,s$^{-1}$ (Table~\ref{tab_atomic}),
and $-75 \pm 10$\,km\,s$^{-1}$ (Table~\ref{TableMolec}), respectively.
These values are less
negative than those estimated in \cite{martini} ($-180$~\kms) and
\cite{tyl} ($-160$~\kms), but are still noticeably different from what is expected
from the Galactic rotation in the direction of V4332~Sgr ($\ga -12$~\kms,
heliocentric). Thus there is no way to estimate
the kinematic distance of the object.

 In the present paper, we adopt, if required, a distance of 1.8~kpc
\citep[following][]{tyl}. This is a very uncertain value, but the principal
conclusions drawn in the present paper are independent of the distance.

\section{Analysis of the molecular bands in emission  \label{embands_mod}}

The main question that arises when analysing the
observed molecular bands in emission concerns  mechanism(s) that could produce these
features. One should consider collisional
excitation and/or radiative pumping of excited states of the molecules as a
primary source of the emissions.
As is shown below, 
the observed band profiles can only be reproduced if the rotational 
temperature is assumed to be very low,
typically $\sim$120~K. This implies that the kinetic temperature must also be
close to it in the ambient medium. In these conditions there is
no way to significantly populate the observed upper electronic states by
collisions, because they would require temperatures of a few thousand K,
meaning that the
only possibility that remains is the radiative pumping. Molecules at ground
states absorb radiation photons capable of exciting them to higher electronic
levels. Radiative decays from them produce the observed emissions. The
analysis done in this section assumes this mechanism.

For each molecule considered in the following subsections, we have
constructed a molecular model including rotational levels of the ground
and excited electronic states. Molecular data for the TiO bands were taken from
\cite{Schwenke}.
The source references for VO and ScO are the same as in \cite{KST09}.
Generally, our models of the molecular structure included rotational levels up to
the quantum number $J=65$.

The population of the rotational levels of the ground
state, including spin substates, is assumed to follow the Boltzman law 
at a given rotational temperature.
The upper electronic state (together with its rotational levels) is assumed
to be populated by radiative transitions from the ground state.
The source of exciting radiation
has usually been approximated by a blackbody of 3200\,K (expected effective
temperature of an M5--6\,III star).
In the case of the TiO bands, a synthetic spectrum from a model stellar atmosphere
of T$_{\mathrm{eff}}$=3200\,K and log\,$g=0.0$ has been calculated
\citep[see][for details]{KST09}. Atmospheric models of \cite{PHOENIX} 
were used for this purpose.
Level populations under the above conditions were solved with 
the code RADEX
\citep{RADEX}. Assuming radiative decays of 
the excited electronic state, a synthetic emission can be calulated 
from the derived populations.
Due to the noise of the observed spectrum, both the observed and calculated
spectra have been smoothed out (typically with an FWHM of 2.4\,\AA) before comparing them.

\subsection{TiO  \label{tio_mod}}

\begin{figure}
  \resizebox{\hsize}{!}{\includegraphics{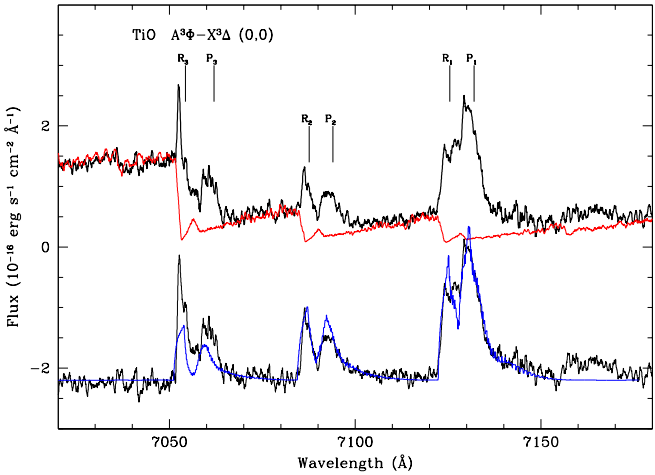}}
  \caption{Top panel: the observed spectrum in the vicinity of the 
TiO $\gamma$ (0,0) band (black line) and a synthetic model spectrum for
$T_{\mathrm{eff}}=3200$~K and $\log g = 0.0$ (red line). Bottom panel: 
result of subtraction of 
the synthetic stellar flux from the observed one (black line) and the model
emission calculated as explained in the text (blue line).
}
  \label{FigTiO}
\end{figure}

Numerous bands of titanium oxide are observed in our spectrum.
Usually narrow emission 
components at the positions of the molecular band-heads 
are superimposed on broad and deep absorption bands of the photospheric
spectrum. The most prominent
examples are three emission components of the TiO $\gamma$ (0,0) band at 7052,
7086, and 7129\,\AA.
We made a model analysis of these three emission components of 
the TiO $\gamma$ (0,0) band.
The results are presented in Fig.~\ref{FigTiO}.

As can be seen from the figure, the emission components 
overlay a broad absorption photospheric component
of the same band with a relatively sharp blue edge at the position of the
F$_3$--F$_3$ emission component at $\sim$7050~\AA. Therefore, for the purpose of the present
analysis of the emission profile, we subtracted a synthetic stellar
spectrum, calculated with $T_{\rm{eff}}=3200$\,K and $\log g=0.0$,
from the observed one (upper part of Fig.~\ref{FigTiO}). 
The result is shown in the lower part of
Fig.~\ref{FigTiO}.

The very first conclusion from the model analysis is that
reliable fits to the individual emission components can be obtained only if the
rotational temperature is assumed to be 120$\pm$20\,K. For an increasing
temperature, the red tail of each component becomes
more pronounced and more extended toward longer wavelengths.
 
Another conclusion can be drawn from the observed flux ratio, which is
2:1:3 for the F$_3$--F$_3$, F$_2$--F$_2$, and F$_1$--F$_1$ components, respectively.
Since the upper substates of each component are connected with the
lower substates only by main branches (satellite ones are very weak), it can
be assumed that all the three emission components are uncoupled transitions.
Then, differences in the fluxes can be caused by different
populations of the lower substates and/or by differences in the flux pumping
the upper levels. Assuming that the lower levels are populated according
to the Boltzman distribution with a temperature of 120\,K, we find that
the ratio of the F$_1$--F$_1$ and F$_2$--F$_2$ fluxes is consistent with the optically thin
limit. 
Effectively, the lowest rotational levels of the F$_3$ substate 
(with energies from 400 to 520 cm$^{-1}$) are about 2 times 
less populated than the lowest rotational levels of the F$_2$ substate,
and 4 times less than those of the lowest F$_1$ substate.
If the pumping flux is similar for the three transitions, 
the expected value of the above flux ratio is therefore 1:2:4.
The observed F$_3$--F$_3$ component is thus significantly enhanced.
An assumption of optically thick
emissions would only flatter this ratio.
The enhanced emission of the F$_3$--F$_3$ component could be explained if the flux 
illuminating the emitting area in this transition were significantly
higher than the fluxes pumping the remaining two. As can be seen from
Fig.~\ref{FigTiO} (upper part) the F$_3$--F$_3$ component (at $\sim 7050$\,\AA) is
situated on the strong blue edge of the photospheric absorption band. Thus,
if the emitting region escapes from the source of the photospheric
radiation,
the edge of the photospheric band is  redshifted and a significant
enhancement of the pumping radiation is then available for the
F$_3$--F$_3$ component compared to the  F$_1$--F$_1$ and F$_2$--F$_2$ components.
Our modelling shows that the observed flux ratio can be satisfactorily
reproduced if the photospheric spectrum, as seen by the emitting region, is
redshifted by at least 25\,km\,s$^{-1}$.
The resulting emission profiles, calculated for a relative velocity between
the photospheric source and the emitting gas of 40\,km\,s$^{-1}$, 
are shown in the lower part 
of Fig.~\ref{FigTiO}.

\subsection{ScO and YO}

\begin{figure}
  \resizebox{\hsize}{!}{\includegraphics{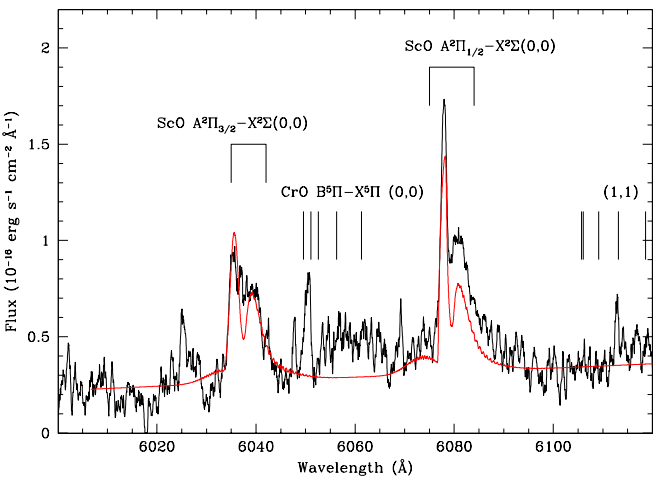}}
  \caption{The observed spectrum around the ScO A$^2\Sigma$--X$^2\Sigma^{+}$ band 
(black line).
  The probable identification of the CrO B$^{5}\Pi$--X$^{5}\Pi$ (0,0) and
(1,1) bands is also marked. 
  A model emission profile of ScO is shown in red. }
  \label{FigScO}
\end{figure}

In the observed spectrum, both ScO A$^2\Pi$--X$^2\Sigma^{+}$ (0,0) 
and YO A$^2\Pi$--X$^2\Sigma^{+}$ (0,0) bands are present in emission.
The rotational contours of ScO sub-bands, 
A$^2\Pi_{3/2}$--X$^2\Sigma^{+}$ and A$^2\Pi_{1/2}$--X$^2\Sigma^{+}$, are shown in
Fig.~\ref{FigScO}. A grid of theoretical profiles have been synthesized
for a range of rotational temperatures. The best fit to the observed spectra
is obtained for a rotational temperature of 120$\pm$20\,K and
a radial velocity of $-$75$\pm$10\,km\,s$^{-1}$ (Fig.~\ref{FigScO}).

Emission of yttrium oxide A$^2\Pi_{1/2}$--X$^2\Sigma^{+}$ (0,0) at 6133\,\AA\ is 
rather weak and barely recognizable. The second
component A$^2\Pi_{3/2}$--X$^2\Sigma^{+}$ (0,0) of YO is not seen in the spectrum.

\subsection{VO}

\begin{figure}
  \resizebox{\hsize}{!}{\includegraphics{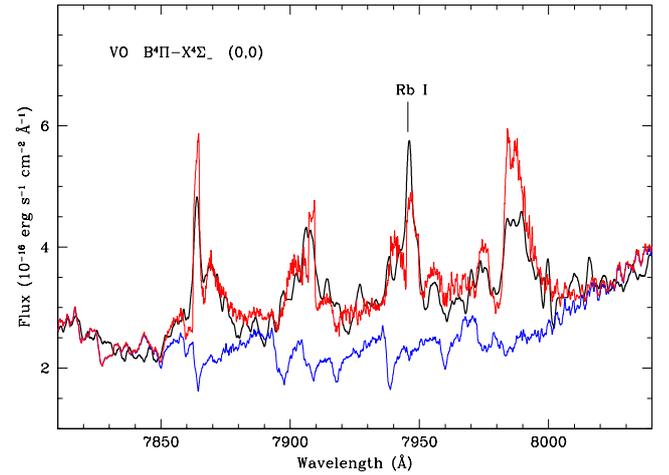}}
  \caption{The observed spectrum near the emission band of VO B$^3\Pi$--X$^4\Sigma_{-}$
  (0,0) (in black). A spectrum of M7~III (HD~18242) star is overplotted in blue.
  The sum of the M7~III spectrum and of the synthesized
  rotational contour is shown in red. See text for details of the model.}
  \label{FigVO}
\end{figure}

Vanadium oxide B$^4\Pi$--X$^4\Sigma^{-}$ (in short B--X) 
(0,0) and (1,0) bands are seen as prominent emission features
(see Fig.~\ref{fig_spec} and Table~\ref{TableMolec}). 
The most intense B--X (0,0) band is shown in detail in Figure \ref{FigVO}.
The band consists of four sub-bands formed by the main branches.
A model fit to the data suggests a rotational temperature 
of $\sim$100~K, in agreement with the rotational temperature 
derived above for the TiO bands.

The observed spectrum around the VO B--X (0,0) band has a relatively high 
S/N. Comparing this part of the spectrum with the UVES
spectrum of HD~18242 (M7~III), we can estimate the radial
velocity of the photospheric spectrum. 
From the sharp absorption feature (probably due to VO in the photosphere) 
at 7960.4\,\AA, as well as the nearby TiO band head at 7818.6\,\AA, 
we estimate the heliocentric stellar velocity to be 
$-$60$\pm$20 km\,s$^{-1}$. 

A final fit was obtained by summing the predicted rotational countour
of the VO~B--X band and the spectrum of HD~18242.
The best fit (shown in red in Fig.~\ref{FigVO}) was obtained
for a velocity of emitting gas of
$-$75$\pm$5\,km\,s$^{-1}$.
The observed spectrum also shows the VO C$^4\Sigma^{-}$--X$^4\Sigma^{-}$
(1,0) and (0,0) bands around 5474\,\AA\ and 5736\,\AA, respectively.
They appear as narrow emission features, which are too weak and too noisy
to be analysed in more detail.

\subsection{CrO}

After having identified the atomic and molecular emission features in the
observed spectrum,
we are left with a few broad ($\sim 20$\,\AA) 
emission-like features in the spectral
range between 5800 and 6800\,\AA. 
We postulate that they come from the electronic band system B$^{5}\Pi$--X$^{5}\Pi$ 
 of CrO observed in emission, as shown in Fig~\ref{fig_spec}. 
Details of the proposed identification can be found in
Table~\ref{TableMolec}.

\begin{figure}
  \resizebox{\hsize}{!}{\includegraphics{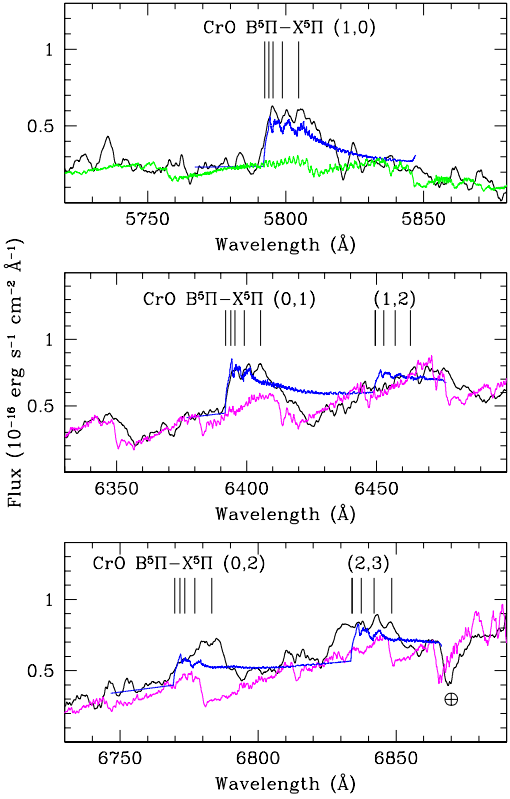}}
  \caption{
Detailed view of CrO B$^5\Pi$--X$^5\Pi$ bands in emission. The observed spectra
are plotted with a black line, while theoretical contours of the CrO bands 
are shown as a blue line.
An M7~III spectrum is displayed in magenta in
the middle and bottom panels. A synthetic spectrum is shown in green in the top
panel. }
  \label{FigCrO}
\end{figure}

We have identified five vibronic bands of CrO, i.e. (1,0), (0,0),
(0,1), (0,2), and (2,3), shown in Figs.~\ref{FigCrO} and 
\ref{FigScO}. The expected positions of the (1,2) and (1,1) bands are also marked
in the figures.
The broad emission profiles are due to overlapping of five sub-bands
formed by main branches. 

Relative intensities of the emission bands with
a common upper vibrational state can be compared with theoretical
predictions. For this purpose we used Franck-Condon
factors computed by \cite{Reddy}. We find that the predicted ratios
are fairly consistent with the observed fluxes. 
The ratios of the emission fluxes of the (1,0):(1,1):(1:2) bands are predicted
to be  3.0:0.1:2.0,
while the observed ones are  7:0:$\la$1, respectively.
The observed ratios of
the (0,0):(0,1):(0:2) bands
are 3.5:4.8:3.4 compared with the theoretical
predictions of  3:4:2,
respectively.

We synthesized the expected contours of
the CrO emission features. A line list from 
the laboratory work of \cite{Hocking} was used. Rotational
intensities were computed consistently from diagonalization of Hamiltonians
defined by \cite{Hocking} 
and using the standard approach \citep[see e.g.][]{Hougen}.
The computed strenghts were checked to reproduce
analytical formulas of rotational intensities of \cite{Kovacs}.
The molecular structure included in the model is rather complex
and consists of the four lowest vibrational levels of
the ground state and the three lowest vibrational levels of the upper
electronic state. Each vibrational level is composed of five
substates and each substate is composed of lower rotational
levels up to $J=65$, in order to include levels with energies up to 2400
cm$^{-1}$.
All the radiative transitions between the listed levels are included.

To reproduce the observed shapes of the rotational contours, 
a rotational temperature of the ground state of 450~K was assumed.
This relatively high rotational temperature (compared to that derived from
the above analysis of other molecules) is needed to explain
formation of all the five band heads seen in the observed spectrum, as well
as their trapezoidal shapes. The exciting
radiation was simulated by a blackbody of 3200\,K.
The obtained theoretical contours are shown in Fig.~\ref{FigCrO}.
They have been shifted
by --70~\kms\ to fit the observed features.

Because of the reasonable consistency between the
observed and predicted 
wavelength positions, rotational contours
(as expected from environment of low kinetic temperature),
intensity ratios, and the
presence of all the expected bands, we consider the identification
of the CrO B--X system in the spectrum of V4332~Sgr as reliable.
CrO was said to be seen
in the atmosphere of $\beta$~Pegasi (spectral class M2)
by \cite{Davis} on the basis of band-head mesurements. Later, on this
finding was questioned by \cite{jorgen}.
Our identification is the first detection of CrO {\it in emission} in any
astrophysical environment.

\section{Origin of the emission features \label{em_origin}}

 Two important conclusions have been drawn from our model analysis of the observed
molecular bands in emission in Sect.~\ref{embands_mod}. First, 
these features are produced by radiative pumping.
Second, the emitting gas is escaping from the source of the pumping radiation
(Sect.~\ref{tio_mod}). 

If the molecular emission bands are formed by radiative pumping then it is
natural to suppose that the atomic emission lines are also produced in the same
mechanism. All the atomic lines observed in our spectrum in
emission are from resonance transitions. 
In this case it is difficult, if not impossible, 
to avoid radiative pumping 
in the atomic lines, if this mechanism is effective for the molecular bands;
both photoabsorption cross sections and the expected abundances are larger for
elements like Na and K than for molecules like TiO, VO, or ScO. 

As can be seen from Table~\ref{tab_atomic}, the observed ratio of the \NaI\,D
lines (5890/5896~\AA) is close to 1.0. The same ratio for the \KI\ lines
(7665/7699~\AA) is $\sim$1.2. In an optically thin case both
ratios are expected to be 2.0. 
The fact that the line ratio in the \NaI\ and \KI\ doublets is observed to be
close to 1.0, can be very easily and naturally explained within the
hypothesis of radiative pumping. The line intensity in this mechanism is
limited by the flux of the pumping spectrum. Since the number of photons available for each
line of a given doublet is practically the same in the pumping spectrum
(assumed to be stellar-like), already for a moderate
thickness (of a few) in the lines, the line ratio must approach 1.0.

In terms of intensities, the \KI\ doublet dominates the spectrum in the
observed spectral range. It is $\sim$25 times stronger that
the \NaI\ doublet. The high
\KI/\NaI\ flux ratio is 
rather unusual. In most circumstellar environments, when both doublets
are observed, the \NaI\ lines are stronger thanks to the $\sim$16
times higher abundance. (The transition
probabilities for both doublets are very close.) The
relative \KI\ and \NaI\ line strengths observed in V4332~Sgr can be
easily explained, if they are assumed to be pumped by radiation. As discussed
above, the observed line ratios indicate that the lines absorb almost all the
available pumping radiation. Therefore the observed \KI/\NaI\ flux ratio is
expected to reflect the spectral energy distribution of the underlying
photospheric source. The flux ratio of the observed photospheric continuum
near the \KI\ and \NaI\ doublets can be estimated from the photometry in the
$I_C$ and $V$ bands, respectively (effective wavelengths of these
photometric bands are close to the wavelengths of the doublets). By adopting
$I_C = 15.1$ and $V = 19.7$ (see Sect.~\ref{sed}) and using the standard
Vega flux calibration, one gets
$\lambda F_{\lambda}(7680\,\AA)/\lambda F_{\lambda}(5890\,\AA) \simeq 28$.
Another way to estimate the above ratio is to measure the continuum from our
Subaru spectrum. Near the \NaI\ doublet, the observed contiunnum is very faint 
but it can be estimated as $F_{\lambda}(5890\AA) \simeq 1 \times
10^{-17}$\,erg\,s$^{-1}$\,cm$^{-2}$\,\AA$^{-1}$. In the
vicinity of the \KI\ doublet, we measure
$F_{\lambda}(7680\AA) \simeq 1.7 \times 10^{-16}$\,erg\,s$^{-1}$\,cm$^{-2}$\,\AA$^{-1}$. 
This results in
$\lambda F_{\lambda}(7680\,\AA)/\lambda F_{\lambda}(5890\,\AA) \simeq 22$.
Both estimates are in excellent agreement with the observed flux ratio of the \KI\
and \NaI\ doublets, confirming our hypothesis that the emission lines
observed in the spectrum of V4332~Sgr are indeed pumped by the photospheric
radiation of the central source.

 As can be seen from Table~\ref{tab_atomic},
the EWs of the \CaI\ and \RbI\ lines are almost two orders of
magnitude lower that those of
the \KI\ lines. This implies that the former
lines are optically thin; i.e., they absorb only a small part of the
available photospheric photons. This statement is confirmed by
the expected low abundance of Rb and the
low transition probability in the \CaI\ (intercombination) line.
The difference in the optical thickness is probably also
the reason for the observed
differences in the profiles, namely that the \CaI\ and \RbI\ lines are
significantly narrower compared to those of \KI\ and \NaI\ (see column 6 in
Table~\ref{tab_atomic}).

We have shown that the spectral shape of the observed
photospheric spectrum of M5-6~III can explain both the shape of the
TiO~$\gamma$~(0,0) band (Sect.~\ref{tio_mod}) and
the flux ratio of the \KI/\NaI\ doublets.
However, in the
observed spectrum we do not see any absorption features, which could be
interpreted as from photons absorbed in the pumping mechanism. This
shows that we do not see any absorbing matter along the line of sight
and implies that the matter seen in the emission features is not distributed 
spherically around the continuum source. The observed continuum 
is also much too faint to be able to pump 
emission features as strong as observed. As can be seen from
Fig.~\ref{fig_spec}, peak intensities in some molecular features and
particulary in the atomic lines are by large factors higher than the
intensity of the underlying continuum. The observed EWs of the
\KI\ lines (see Table~\ref{tab_atomic}) imply that 
if they had been excited by the observed continuum in their
vicinity, they would have had to absorb all the continuum photons from a
spectral region as large as $\sim$550~\AA. This would be possible only if
the absorbing matter had been moving with a velocity gradient, between
slowest and fastest regions, as large as $\sim$20\,000~\kms, which is two
orders of magnitude higher than the
velocities observed in V4332~Sgr during its eruption \citep{martini}. Therefore
we conclude that the photospheric continuum responsible for pumping the
observed emission features is much stronger than the observed one. The above
discussion of the \KI\ lines suggests that we observe $\la$1\% of the
flux emitted by the central object. In Sect.~\ref{sed} we further argue
that the central object in V4332~Sgr is mostly hidden to us by an opaque disc 
seen almost edge-on.

\section{Spectral energy distribution}\label{sed}

The SED of V4332~Sgr was already investigated by
\cite{tyl}. It was based on photometric data obtained in 2003 and covering
photometric bands from $B$ to $M$. \cite{tyl} point out that apart from
the stellar-like component dominating in the optical and near-IR, the object
displayed a large IR-excess seen in the $KLM$ bands. When interpreted as a
single blackbody the IR component had an effective temperature of 
$\sim$750~K and a luminosity $\sim$15 times higher than that of the optical component.
Another possibility mentioned in \cite{tyl} was an accretion disc
$\sim$25 times more luminous than the central star.

\begin{figure}
\centering
   \includegraphics[height=\hsize,angle=270]{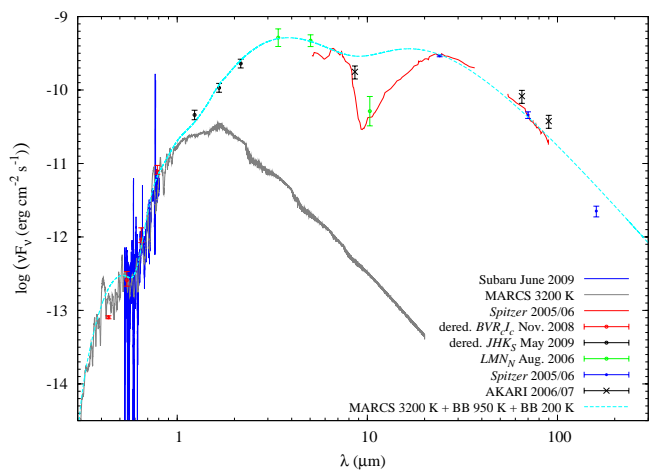}
   \caption{Spectral energy distribution for V4332 Sgr. Observational
     points are taken from different sources as listed in the
     text.  
     Our attempt to fit the observed SED is shown with a dashed
     line.}  
 \label{fig_sed}
\end{figure}

Figure \ref{fig_sed} displays an SED of V4332~Sgr
derived from various observations obtained between April~2005 and
May~2009. The source of the data are as follows:
\begin{itemize}
\item Optical photometric measurements were taken from the web page of
  V. Goranskij\footnote{http://jet.sao.ru/$\sim$goray/v4332sgr.ne3}. These
  are combined data from different observatories. We interpolated the
  $BVR_CI_C$ magnitutes for the date of the Subaru/HDS observations
  and adopted $B$=21.4, $V$=19.7, $R_C$=17.9, $I_C$=15.1 mag. The
  error bars shown in Fig.~\ref{fig_sed} for the optical magnitudes
  represent a scatter in the photometric measurements around  the date
  of the Subaru/HDS observations.
\item The $JHK_S$ data are from the observations reported on in Sect.~\ref{tcs}. 
\item IR measurements in the $LMN_n$ bands were taken from \cite{lynch}, and they
correspond to observations carried out on 7 August 2006.
\item Far-IR data were taken from \cite{banerSpitzer}. These are
  {\it Spitzer}/MIPS fluxes at 24~$\mu$m, 70~$\mu$m, and 160~$\mu$m 
averaged from measurements obtained on 15 October 2005
  and 2 November 2006. 
\item  AKARI measurmements were taken form the first releases of the FIS Bright
Source Catalogue \citep{fis} and IRC Point Source Catalogue \citep{irc},
which are based on observations from the AKARI All-Sky Survey performed
in 2006--2007. For the AKARI Infrared Camera (IRC), the source was
detected only in the 9~$\mu$m band with flux density of 507.7~mJy.
Reliable measurements with the Far-Infrared Surveyor instrument (FIS)
are available only for the 65~$\mu$m and 90~$\mu$m bands with density
fluxes of $F_{65}$=1.79~Jy and $F_{90}$=1.13~Jy. There are also data for
the FIS bands at 140 and 160~$\mu$m ($F_{140}$=86.9~mJy and
$F_{160}$=29.9~mJy), but the measurements are flagged as unreliable 
(too weak a source at those bands). The total (absolute and relative)
calibration uncertainty of the measurements in all the AKARI bands is
20\%\ \citep[3$\sigma$;][]{fis,irc}.
\item In Fig.~\ref{fig_sed} we also display spectra of V4332 Sgr in wide
  spectral ranges. The optical spectra are the Subaru/HDS spectrum
  reported in this paper and a synthetic spectrum from the MARCS grid \citep{marcs}
  with $T_{\rm eff}=3200$~K, solar metalicity, and $\log g=0.0$; the last was
  rescaled to fit the fluxes of the Subaru/HDS data. The {\it Spitzer} spectra 
are also shown. They were acquired
  on 18 April 2005 and 27 September 2005.
\end{itemize}

Vega fluxes were used as zero points when converting magnitudes to
flux-density units. 
The optical spectra and the $BVR_CI_CJHK_S$ data points shown in
Fig.~\ref{fig_sed} are corrected for interstellar reddening with
$E_{B-V}=0.32$ and $R_V=3.1$. 

In the following discussion we assume that the object did not vary
significantly between 2005 and 2009. 
This is not entirely justfied since, as noted in Sect.~\ref{intro}, the
object faded in the optical in 2006; however, it
seems that the
overall SED has remained more or less unchanged over a few recent years.
When compared with the
photometric data in 2003 \citep{tyl}, the object is now $\sim$1.5~mag
fainter in $BVR$ but $\sim$1.0~mag brighter in $JHK$. \cite{banerSpitzer}
note that the object remained constant over a year in their far-IR
observations in 2005--2006.

Similar to \cite{tyl}, the SED in Fig.~\ref{fig_sed} displays a huge
IR-excess. However, thanks to the far-IR data, we can now conclude that it
cannot be represented by a single temperature component. We have attempted
to reproduce it in Fig.~\ref{fig_sed} with a sum of two blackbodies of 950~K 
and 200~K, but it would
be more reasonable to conclude that this is a multi-temperature component 
with temperatures ranging between these two values. This result, combined
with the fact that the IR component has remained stable since at least 2003, 
strongly suggests that we see a multi-temperature Keplerian disc 
of matter surrounding the central object resembling an M5--6 giant.

In the simple fit to the observed points in Fig.~\ref{fig_sed} the two IR
components, i.e. blackbodies of 950~K and 200~K, are $\sim$30 and
$\sim$20 times more luminous, respectively, than the stellar component of 
$T_{\rm eff}=3200$~K.  Assuming spherical geometry of the components, the two
infrared components would have radii $\sim$65 and $\sim$1200 times larger than the
stellar one.
If interpreted with a disc-like spectrum, the disc
would have an apparent luminosity $\sim$50 times higher than the central star. 

In principle,
one can consider two possible sources of energy radiated away by a disc,
i.e., reproduction of the central star radiation absorbed by the disc matter
(so-called passive disc) or gravitational energy dissipation by viscous
forces transferring matter from outer disc layers to inner ones and
eventually to the star surface (accretion disc). In both cases, the central
star is expected to be at least as bright as the disc. In the first case
(passive disc), for obvious geometrical reasons the disc can absorb,
hence reradiate, only a part of the stellar luminosity. In the second case
(accretion disc), the accreted matter is expected to release a comparable
energy to what is dissipated in the disc, at the star surface (so-called disc
boundary layer), where the matter has to slow down from the Keplerian
velocity to the velocity of stellar rotation (usually much lower than
Keplerian). In the case of a giant or a supergiant, the boundary-layer braking is
likely to take place below the photosphere. In this case, accretion is expected to
add a luminosity to the intrinsic stellar luminosity, similar to that of the
disc.

The only way to reconcile that the central star in V4332~Sgr is
observed to be $\sim$50 times less luminous that the disc is to assume that the
disc is observed almost edge-on. In this case the disc would block most of
the stellar radiation in the direction to the observer. The
geometrical aspect (inclination close to $90\degr$) means that the observed lumnosity of
the disc is also expected to be significantly lower than the true one.
Therefore we can conclude that the stellar component observed
in the optical only accounts for $\la$1\% of the luminosity of the
central object in V4332~Sgr. This conclusion nicely fits the one drawn from
our analysis of the emission features in Sect.~\ref{em_origin}.

\section{Discussion}

Our analysis done in Sects.~\ref{em_origin} and \ref{sed} has led us to
conclude that V4332~Sgr is now observed as consisting of a central object,
resembling an M5--6 giant, surrounded by a disc of circumstellar matter. 
The inclination
angle of the disc is close to $90\degr$, so that the disc matter obscures the
central star, and we can only observe a tiny part of its radiation. The
situation thus resembles what is observed in IRAS\,18059--3211, known as 
Gomez's Hamburger \citep{ruiz,wood}, whose global SED is
similar to V4332~Sgr.
The central star in Gomez's Hamburger is not directly seen,
and the observed optical spectrum is entirely due to scattering of the central
star radiation on dust grains in the outer edge of the disc. It is very
probable that the same situation is also observed in V4332~Sgr. Polarimetric
observations in the optical would allow verification of this hypothesis.

 Molecular bands in emission have been observed
in the optical spectrum of the luminous red supergiant VY~CMa \citep[][and references
therein]{HER74,WAL01}, as well as of U~Equ \citep{Barnbaum}.
In both cases it has been suggested that the objects are
surrounded by massive circumstellar discs, which obscure the
central stars along the line of sight
\citep{her70,KAS98,Barnbaum}, i.e. similar to what we propose in the present paper
for V4332~Sgr. The existence of a disc obscuring the central star thus seems to
be a condition
favouring observations of molecular bands in emission. Indeed, if we
increased the photospheric spectrum in V4332~Sgr by a factor of 100 (as
suggested in Sects.~\ref{em_origin} and \ref{sed}, if the central star were
unobscured) remaining the fluxes of the emission features unchanged, then we
would have practically no chance to detect any emission molecular band.
The only visible emission features would then be those of \KI\ and \NaI.

As shown in Sect.~\ref{sp_phot}, the spectral
type of the observed photospheric spectrum agrees with the $BVR_cI_c$
photometry corrected for the interstellar extinction.
Thus the scattering responsible for the observed photospheric spectrum
 has no significant effect on its global shape. This can be understood if
the observed spectrum is mainly produced by forward scattering on the outer
disc rim lying in front of the central star, as we suggest for V4332~Sgr. In
this case scattering is dominated by large grains ($a \ga 0.3\,\mu$m), which
have a high value of $g \equiv <{\rm cos}\,\theta>$, i.e. scatter mainly in
forward directions. These grains also have $Q_{\rm sca}$ that is practically
independent of wavelength in the optical \citep[e.g.][]{ld93}.\footnote{see
also http://www.astro.princeton.edu/$\sim$draine/dust/dust.html}

As discussed in Sect.~\ref{em_origin}, the emission features observed in
V4332~Sgr are most likely pumped by radiation from the central star. One may ask
 where the matter seen in the emission features can be situated. One
possibility is that this could be matter in the atmosphere of the disc.
However, the profiles of the emission lines observed in V4332~Sgr are
single-peaked (see Fig.~\ref{fig1}), while it is
widely believed that emission features observed from an edge-on disc
should be double-peaked (due to the disc rotation). Although this is
true in most cases, there are situations where disc emission lines
can be observed as single-peaked.
Simulations of \citet{windCV,windAGN} show that this happens
when the disc is seen at a high inclination
angle and possesses a wind, which is optically thick in the lines. 
This mechanism could explain the profiles
of the \NaI\ and \KI\ lines, since, as discussed in Sect.~\ref{em_origin},
these lines are most likely optically thick. 
 However, as discussed in the same section, the \CaI\ and \RbI\
lines are optically thin, and yet they
are single-peaked (see Fig.~\ref{fig1}). 
Another problem arises from the observed line wings
extending up to $\sim\pm$250~\kms\ from the line centre (see
Fig.~\ref{fig1}). If formed in a high-inclination disc, they would measure
the rotation velocity of the innermost observable regions of the disc.
 Adopting a distance of V4332~Sgr of $\sim$1.8~kpc
\citep[following][who obtained this value after
assuming that the progenitor was a solar type star]{tyl}, our fit to the
optical photometry (mentioned in Sect.~\ref{sp_phot}) results in an
effective radius of $\sim$5~R$_\odot$ \citep[similarly to the one derived in][for
2003]{tyl}. However, as we now know, the
true central star is probably two orders of magnitude brighter than observed
and thus has a radius that is an order of magnitude larger than the above estimate. 
For a solar-mass star and a radius of $\sim$50~R$_\odot$ one
obtains a Keplerian velocity of $\sim$60~\kms\ at the stellar surface. Since
the central star is obscured and thus the disc regions up to a few stellar
radii are also unseen, there seems to be no way to explain the line wings
extending up to $\sim\pm$250~\kms\ within the idea of the emission lines
arising from the disc. A higher stellar mass would make the situation even
worse. For instance, a 3~M$_\odot$ progenitor would be almost two orders of
magnitude brighter and about 10 times more distant, and the outburst remnant would
be 10 times more extended, lowering the Keplerian velocity at its surface by a
factor of 2.

It thus seems that we are only left with a scenario, in which the emission features
are formed in matter lost by the central object. This idea is in accord with
our findings in Sect.~\ref{tio_mod} that the observed ratio of the
emission components of the TiO~$\gamma$~(0,0) band implies an outflow
velocity of the emitting matter of at least 25~\kms. 
The origin of the outflowing matter could be
matter ejected in the past, e.g. during the 1994 outburst, or a wind flowing out 
at present from the object, or both. The low rotational temperature found from the
molecular bands in emission (Sect.~\ref{embands_mod}) suggests that these
features are formed in a cold medium, thus at significant distances from
the central object. This seems to favour the former possibility of matter
ejected in the past.
The observed widths of the atomic lines at zero intensity (see
Fig.~\ref{fig1}) imply an outflow velocity of up to $\sim 250$~\kms. The line
profiles suggest that the outflow is concentrated along an axis nearly
perpendicular to the line of sight. This could be the axis of the disc.
Detailed modelling of the line profiles would be required to better
constrain the structure and nature of the outflow.

\acknowledgements{We would like to thank the group of support astronomers at
  the Carlos S\'{a}nchez Telescope, especially R. Barrena, for obtaining
  the service photometric observations reported in this paper. 
 This research made use of the NASA/ IPAC Infrared Science Archive, 
 which is operated by the Jet Propulsion Laboratory, California Institute of Technology, 
under contract with the National Aeronautics and Space Administration. 
We acknowledge the use of data from the UVES Paranal Observatory
Project (ESO DDT Programme ID 266.D-5655). 
This research is also partly based on observations with AKARI, 
a JAXA project with the participation of ESA.
The research reported on in this paper has partly been supported by a grant
no. N203~004~32/0448 financed by the Polish Ministery of Sciences and Higher
Education.} 
   
\bibliographystyle{aa}

\begin{thebibliography}{}

\bibitem[Bagnulo et al.(2003)]{uves} 
Bagnulo, S., Jehin, E., Ledoux, C., et al. 
2003, The Messenger, 114, 10 

\bibitem[Banerjee et al.(2003)]{baner3}
Banerjee, D. P. K., Varricatt, W. P, Ashok, N. M., \& Launila, O.
2003, \apj, 598, L31

\bibitem[Banerjee \& Ashok(2004)]{baner4} 
Banerjee, D.~P.~K., \& Ashok, N.~M.\ 2004, \apjl, 604, L57 

\bibitem[Banerjee et al.(2004)]{banerCO} Banerjee, D.~P.~K., 
Varricatt, W.~P., \& Ashok, N.~M.\ 2004, \apjl, 615, L53 

\bibitem[Banerjee et al.(2007)]{banerSpitzer} 
Banerjee, D.~P.~K., Misselt, K.~A., Su, K.~Y.~L., Ashok, N.~M., 
\& Smith, P.~S.\ 2007, \apjl, 666, L25 

\bibitem[Barnbaum et al.(1996)]{Barnbaum}
Barnbaum, C., Omont, A., \& Morris, M. 1996, A\&A, 310, 259

\bibitem[Bohlin et al.(1978)]{bohlin} 
Bohlin, R.~C., Savage, B.~D., \& Drake, J.~F.\ 1978, \apj, 224, 132 

\bibitem[Brand \& Blitz(1993)]{bb93}
Brand, J. \& Blitz, L. 1993, \aap, 275, 67

\bibitem[Buckle et al.(2009)]{harp} 
Buckle, J.~V., et al. 2009, \mnras, 399, 1026 

\bibitem[Davies(1947)]{Davis}
Davis, D. N. 1947, ApJ, 106, 28

\bibitem[Gunn \& Stryker(1983)]{gs} 
Gunn, J.~E., \& Stryker, L.~L.\ 1983, \apjs, 52, 121 

\bibitem[Gustafsson et al.(2008)]{marcs} 
Gustafsson, B., Edvardsson, B., Eriksson, K., et al. 
2008, \aap, 486, 951 

\bibitem[Hamuy et al.(1994)]{ham1} 
Hamuy, M., Suntzeff, N.~B., Heathcote, S.~R., et al.  
1994, \pasp, 106, 566 

\bibitem[Hamuy et al.(1992)]{ham2} 
Hamuy, M., Walker, A.~R., Suntzeff, N.~B., et al. 
1992, \pasp, 104, 533 

\bibitem[Hauschild et al.(1999)]{PHOENIX}
Hauschildt, P. H., Allard, F., Ferguson, J., Baron, E., \& Alexander, D. R.
1999, \apj, 525, 871

\bibitem[Herbig(1970)]{her70}
Herbig, G. H. 1970, \apj, 162, 557

\bibitem[Herbig(1974)]{HER74}
Herbig, G. H. 1974, \apj, 188, 533

\bibitem[Hocking et al.(1980)]{Hocking}
Hocking, W. H., Merer, A. J., Milton, D. J., Jones, W. E., \& Krishnamurty,
G. 1980, Can. J. Phys., 58, 516

\bibitem[Hougen(2001)]{Hougen}
Hougen, J. T. 2001, 'The Calculation of Rotational Energy Levels 
and Rotational Line Intensities in Diatomic Molecules' (version 1.0),
available at: http://physics.nist.gov/DiatomicCalculations [2009, December 9],
National Institute of Standards and Technology, Gaithersburg, MD.

\bibitem[Jorgensen(1996)]{jorgen} Jorgensen, U.~G.\ 1996, 
in Molecules in Astrophysics: Probes \& Processes, IAU Symp. 178 (ed. E. van
Dishoeck), p.441 

\bibitem[Kalberla et al.(2005)]{lab} 
Kalberla, P.~M.~W., Burton, W.~B., Hartmann, D., et al.  
2005, \aap, 440, 775  

\bibitem[Kami{\'n}ski et al.(2009)]{KST09} Kami{\'n}ski, T., 
Schmidt, M., Tylenda, R., Konacki, M., 
\& Gromadzki, M.\ 2009, \apjs, 182, 33 

\bibitem[Kastner \& Weintraub(1998)]{KAS98}
Kastner, J. H., \& Weintraub, D. A. 1998 AJ, 115, 1592

\bibitem[Kataza et al.(2010)]{irc}
Katza, H. et al. 2010, AKARI/IRC All-Sky Survey Point Source Catalogue
Version 1.0 available at
http://www.ir.isas.jaxa.jp/AKARI/Observation/PSC/Public.

\bibitem[Kimeswenger(2006)]{kimes} 
Kimeswenger, S.\ 2006, Astronomische Nachrichten, 327, 44 

\bibitem[Kimeswenger(2007)]{kimesConf} 
Kimeswenger, S.\ 2007, ASP Conf. Ser., 363, 197 

\bibitem[Kovacs(1969)]{Kovacs} Kovacs, I. 1969, Rotational Structure in the Spectra of
Diatomic Molecules (Budapest)

\bibitem[Loar \& Draine(1993)]{ld93}
Loar, A. \& Draine, B. T. 1993, \apj, 402, 441

\bibitem[Lynch et al.(2006)]{lynch} Lynch, D.~K., Russell, 
R.~W., Ford, R., Hammel, H.~B., \& Sitko, M.~L.\ 2006, \iaucirc, 8739, 1 

\bibitem[Martini et al.(1999)]{martini} 
Martini, P., Wagner, R.~M., Tomaney, A., et al.  
1999, \aj, 118, 1034  

\bibitem[Munari \& Zwitter(1997)]{ebv} 
Munari, U., \& Zwitter, T.\ 1997, \aap, 318, 269 

\bibitem[Murray \& Chiang(1997)]{windAGN} 
Murray, N., \& Chiang, J.\ 1997, \apj, 474, 91 

\bibitem[Murray \& Chiang(1996)]{windCV} 
Murray, N., \& Chiang, J.\ 1996, \nat, 382, 789 

\bibitem[Noguchi et al.(2002)]{hds}
Noguchi, K., et al.\ 2002, \pasj, 54, 855 

\bibitem[Phillips(1973)]{Phillips1973}
Phillips, J.G. 1973, \apjs, 26, 313

\bibitem[Reddy et al.(1998)]{Reddy}
Reddy, R. R., Nazeer Ahammed, Y., Rama Gopal, K., Abdul Azeem, P.,
\& Anjaneyulu, S. 1998, Ap\&SS, 262, 223

\bibitem[Richichi et al.(1999)]{mgiants} 
Richichi, A., Fabbroni, L., Ragland, S., \& Scholz, M.\ 1999, \aap, 344, 511 

\bibitem[Ruiz et al.(1987)]{ruiz}
Ruiz, M. T., Blanco, V., Maza, J. et al. 1987, \apj, 316, L21

\bibitem[Rosen(1970)]{Rosen1970} 
Rosen, B., 1970, Spectroscopic Data relative to Diatomic Molecules,
 Pergamon Press.

\bibitem[Sch{\"o}ier et al.(2005)]{LAMDA}
Sch{\"o}ier, F.L., van der Tak, F.F.S., van Dishoeck, E.F., Black, J.H.
2005, A\&A, 432, 369

\bibitem[Schwenke(1998)]{Schwenke}
Schwenke, D. W. 1998, Chemistry and Physics of Molecules 
and Grains in Space, Faraday Discussions No., 109, 321

\bibitem[Soker \& Tylenda(2003)]{st03}
Soker, N. \& Tylenda, R. 2003, \apj, 582, L105

\bibitem[Tylenda et al.(2005)]{tyl} 
Tylenda, R., Crause, L.~A., G{\'o}rny, S.~K., \& Schmidt, M.~R.\ 2005, \aap, 439, 651 

\bibitem[Tylenda \& Soker(2006)]{ts06}
Tylenda, R. \& Soker, N. 2006, \aap, 451, 223

\bibitem[van der Tak et al.(2007)]{RADEX}
{{van der Tak}, F.~F.~S., {Black}, J.~H., {Sch{\"o}ier}, F.~L., 
{Jansen}, D.~J., and {van Dishoeck}, E.~F.} 2007 A\&A, 468, 627

\bibitem[Wallerstein \& Gonzalez(2001)]{WAL01}
Wallerstein, G., \& Gonzalez, G. 2001, PASP, 113, 954
  
\bibitem[Willmarth \& Barnes(1994)]{irafech}
Willmarth, D. \& Barnes, J. 1994, A User's Guide to Reducing Echelle
Spectra With {\it IRAF} (Tucson: NOAO), http://iraf.net/irafdocs/ech/ 

\bibitem[Wilson et al.(2009)]{tools} 
Wilson, T.~L., Rohlfs, K., H{\"u}ttemeister, S.\ 2009, Tools of Radio
Astronomy (Springer, Berlin) 

\bibitem[Wood et al.(2008)]{wood}
Wood, K., Whithney, B. A., Robitaille, T., \& Draine, B. T.
2008, \apj, 688, 1118

\bibitem[Yamamura et al.(2010)]{fis}
Yamamura, I. et al. 2010, AKARI/FIS All-Sky Survey Bright Source
Catalogue Version 1.0, available at
http://www.ir.isas.jaxa.jp/AKARI/Observation/PSC/Public.

\end{thebibliography}

\end{document}